\documentclass[pdflatex,sn-basic,Numbered,iicol]{sn-jnl}
\usepackage{amsmath,amssymb,amsfonts}
\usepackage{graphicx}
\graphicspath{{figures/}{./}}
\usepackage{xcolor}
\usepackage{manyfoot}
\DeclareNewFootnote{A}[gobble]
\usepackage{booktabs}
\usepackage{array}
\hypersetup{bookmarksdepth=2,
  pdftitle={Reading the Data Back: Enriching Variable-Level Metadata for Model-Data Consistency Checks},
  pdfauthor={Eryk Kulikowski},
  pdfkeywords={research data repositories, DDI-CDI, variable-level metadata, metadata application profiles, provenance, analysis fingerprints}}

\begin{document}

\title[Reading the Data Back]{Reading the Data Back: Enriching Variable-Level Metadata for Model-Data Consistency Checks}

\author*[1]{\fnm{Eryk} \sur{Kulikowski}}\email{eryk.kulikowski@kuleuven.be}
\affil*[1]{\orgdiv{LIBIS}, \orgname{KU Leuven}, \orgaddress{\city{Leuven}, \country{Belgium}}}
\artnote{Author ORCID: \url{https://orcid.org/0000-0002-9967-9732}.}

\abstract{\textbf{Purpose}\enspace Research data repositories often lack variable-level metadata. Model-choice screening, checking whether a reported model suits the values of its
outcome variable, also requires summaries of those values and links to the analyses that use them. We
ask how repositories can represent and acquire them as metadata with evidence and review histories.

\textbf{Methods}\enspace We propose a metadata application profile compatible with the Data Documentation Initiative
Cross-Domain Integration (DDI-CDI) model, linking versioned variables and empirical profiles to
reported analyses, estimators, and analysis roles. Extraction provenance and review decisions are
recorded separately. We evaluate streaming data profiling, deterministic extraction from Stata and R
code, and language-model extraction of analysis records from papers.

\textbf{Results}\enspace A worked export of one replication package illustrates the profile: it passes shape and interchange
checks and answers four queries, one of which returns the analyses that raise no alert. Screening 4{,}868 replication datasets from
six political-science journals links outcomes to profiled columns in 1{,}440 deposits and flags 965,
including 14 of the 21 deposits with a hand-verified linear model on a count, proportion, binary, or
ordinal outcome. AI-assisted adjudication yields 55\% precision on a near-balanced sample of 119
count and proportion candidates; screening rules were developed on the same corpus.

\textbf{Conclusion}\enspace The profile makes analysis--variable relationships queryable while preserving evidence and review
history; screening yields review candidates with context. Code, metadata, and
measurements are released.
}

\maketitle

\section{Introduction}
\label{sec:intro}

\begin{table*}[t]
\caption{The worked example and the two evaluations, each read on its own.}
\label{tab:map}
\small
\setlength{\tabcolsep}{4pt}
\begin{tabular*}{\textwidth}{@{\extracolsep{\fill}}>{\raggedright\arraybackslash}p{0.27\textwidth}>{\raggedright\arraybackslash}p{0.24\textwidth}>{\raggedright\arraybackslash}p{0.43\textwidth}@{}}
\toprule
component & material & what it establishes \\
\midrule
application-profile example (Section~\ref{sec:profile}) & one deposit, four analyses & representation, queries, validation, interchange \\
repository census (Section~\ref{sec:atscale}) & 4{,}868 deposits; benchmark recovery & coverage and screening yield from data and code \\
fingerprint experiment (Section~\ref{sec:textroute}) & 37 benchmark cases & outcome-link and estimator-family recovery from papers, with and without code \\
\bottomrule
\end{tabular*}
\end{table*}

Research data repositories hold many datasets whose variables are described thinly or not at all.
Generators that read the deposited tabular files can fill part of that gap; the one we extend emits
Data Documentation Initiative Cross-Domain Integration (DDI-CDI) metadata with a datatype for every
column~\citep{rdmint,ddicdi}. To check a reported model against its own data, a repository also
needs to know what values a variable actually takes and which analyses used it. This paper adds both: empirical profiles of the observed values, and explicit links
from variables to reported analyses, each carrying its evidence and its review state.

Take one replication package from the 40 cases we verify below (Section~\ref{sec:testbed}), a
study of hate crimes in India. Its data file has a column of hate-crime counts, \texttt{hcam1}.
Nothing in the deposit's metadata says which reported analysis uses that column, that it is the
response of the paper's headline regression, that the paper's own robustness check refits the same
column with a count model, who asserted any of this, from which evidence, or whether anyone has
reviewed it. The dataset and the analyses that used it meet only in the head of a reader who has
both open at once. Reading the data back means preserving those relationships, so that the next
reader can retrieve the headline result together with its robustness check, and inspect both.

Once that link is metadata, questions that today require reading paper and code side by side become
queries over the archive: which analyses use this variable as their response, which of them carry the
headline claim, which observations were profiled, and which mappings someone has reviewed. This paper
builds that metadata from linked deposits and papers and specifies how each claim attaches to a
versioned variable together with its evidence, its generation method, and its review state
(Section~\ref{sec:profile}). We test the representation with executable queries and interchange
checks. We test the three ways of filling it, from data, code, and papers, with extraction and
coverage measurements. Model-choice screening
is the use case throughout, because it needs both kinds of evidence, a property of the data and a fact
about a reported analysis, and because an established literature supplies the review rubric.

In 1988 Gary King showed that ``hundreds of important political science studies'' modeled event
counts with ordinary least squares~\citep{king}. The fractional-response literature makes a similar
case for bounded proportions~\citep{papke,warton}, while the linear probability model (LPM) for binary
outcomes~\citep{angrist} and ordinal scales treated as continuous~\citep{long} remain debated rather
than settled. The archives that make such choices checkable now exist: a large share of quantitative
political science deposits its data and analysis code in replication archives, prominently Harvard
Dataverse~\citep{king2007,crosas}. The instances have not been examined systematically: to our
knowledge no labeled corpus of published analyses exists for this class of estimator--outcome
combination, and per-paper model-family critique is a methodologists' genre rather than a routine
check (Section~\ref{sec:testbed}). Among the 40 cases verified below are linear models on logged
battle deaths, on vote shares and turnout rates bounded in $[0,1]$, and a Gaussian ANOVA on species
counts. In the opening example, a reconstruction of the hate-crime regression produces negative fitted counts
for 29\% of its observations, and the paper reports quasi-Poisson robustness checks for the same
outcome~\citep{basu}. Both facts matter. A linear estimator can estimate an average treatment effect
under suitable design conditions~\citep{angrist,lin2013}, and a negative fitted value cannot be a
count's conditional mean at that observation. Recording the headline and the robustness analyses as
separate, linked records keeps that context next to the diagnostic, so what the repository holds is a
review candidate with evidence and a scope, not a verdict on a variable or a paper.

\paragraph{What this paper builds, and what it runs}
Three terms recur. The \emph{application profile} is the metadata specification proposed here. An
\emph{empirical profile} is one variable's summary of its observed values. An \emph{analysis
fingerprint} is the set of analysis records extracted from one paper. The representation and the
ways of filling it are evaluated separately (Table~\ref{tab:map}). The example export in
Section~\ref{sec:profile} is a worked illustration on one deposit; its released checks confirm that the
shipped graph, shapes, and queries behave as the text says. The empirical questions, coverage and
yield, are answered by the census and the fingerprint experiment. On the data side, the
deposited dataset is profiled variable by variable into DDI-CDI metadata plus an inferred operational
class (Section~\ref{sec:measured}). On the analysis side, an analysis fingerprint is distilled from
the linked paper: a record for each reported analysis the prompt finds, with its stated estimator
family, its modeled variable
mapped to a dataset column, and whether it is the headline claim or a side regression
(Section~\ref{sec:textroute}). Deposited code is a second source for the analysis side, read either
by an auditable keyword mapping over Stata and R estimator commands or by adding the scripts to the
fingerprint call (Sections~\ref{sec:atscale} and~\ref{sec:textroute}). Once both sides are metadata,
the check is a query. A flag marks an estimator--outcome combination that a declared rule selects for
review, here a linear estimator on selected non-continuous outcome classes; it does not reconstruct the
authors' full statistical assumptions. A flag is a place to look twice, at the analysis and at our own
metadata alike: it can mark a model choice worth revisiting or a variable that the profiler or the
extractor has misread, and either is worth a reviewer's attention. The example export exercises the
application profile itself. The census measures only the join underneath it, from estimator evidence
in code to the profiled response column; it does not materialize the full profile or extract
analysis roles across the archive.

\paragraph{Contributions}
\begin{enumerate}
\item \textbf{A specified and tested application profile}: versioned DDI-CDI
\texttt{InstanceVariable} identities linked to empirical profiles, reported analyses, analysis-role
and variable-use assertions, and separate extraction and review records. An example export tests
four queries, RDF (Resource Description Framework) interchange, and distinct core and extension
constraints
(Section~\ref{sec:profile}).
\item \textbf{Evidence acquisition from data, code, and papers}: a streaming empirical profiler, a
deterministic estimator extractor, and per-analysis fingerprints with outcome links and analysis
roles. Their evaluation separates variable classification, extraction fidelity, and abstention
(Sections~\ref{sec:measured} and~\ref{sec:textroute}).
\item \textbf{A repository-scale screening case study}: a 40-case source-checked benchmark and a census
of 4{,}868 replication datasets quantify recoverable links, flag volume, adjudicated precision,
and fitted-value diagnostics. The released evidence identifies where missing code, derived variables,
and ambiguous profiles prevent the metadata join or change its interpretation
(Sections~\ref{sec:testbed} and~\ref{sec:atscale}).
\end{enumerate}

\section{Related work}
\label{sec:related}

The paper draws on three lines of work: variable-level metadata and profiling, provenance-bearing
research objects and assertions, and the extraction or checking of reported analyses. The profile
supplies the relationships between them that analysis-scoped queries need.

\paragraph{Variable metadata and empirical profiles}
DDI-CDI supplies identities and descriptions for variables, value domains, and processes~\citep{ddicdi}.
The open-source generator we extend emits this metadata from tabular files~\citep{rdmint}.
Machine-actionable metadata is an explicit aim of FAIR~\citep{fair}, and systems such as Deequ
compute and validate data properties at scale~\citep{deequ}. Our empirical profiles follow this approach, adding observed support and shape statistics and an operational class used by the screen.
The distinction is one of scope: a value domain describes allowable values; an empirical profile
describes particular observations and the rules used to summarize them. The inferred class may
change when the sample changes even though the allowable domain does not.

\paragraph{Research objects, provenance, and assertions}
RO-Crate packages research artifacts and their contextual relationships in JSON-LD~\citep{rocrate}.
Nanopublications make individual assertions and their provenance independently identifiable~\citep{nanopub}.
PROV-O describes generation activities, agents, and revisions~\citep{provo}; Web Annotation identifies
supporting regions of source documents~\citep{annotation}. These are the general mechanisms for
packaging and attributing claims. Our profile specializes them to one kind of claim, that a reported
analysis uses a versioned variable in a given role, under a given data scope, with given extraction
and review records; the contribution is that combination of relationships and constraints, not a new
mechanism. The prototype uses DDI-CDI, PROV-O, and Web Annotation; mapping its assertions into
RO-Crate or nanopublication packages would be a further integration.

\paragraph{Analysis evidence and computational fingerprints}
A domain-stripped computational fingerprint represents a paper's computational structure independently
of its topic. Earlier work evaluates it for cross-domain solution import~\citep{p1}, and a companion
preprint studies replication prediction~\citep{p3}. Both operate on papers alone. Here the
representation gains per-analysis records that name the reported estimator, the outcome, the analysis
role, and the link to a deposited variable. Section~\ref{sec:textroute} tests those links from paper
text and metadata, then with code added; the role labels are extracted but not validated.

\paragraph{Reproduction and reanalysis from deposits}
Large-scale processing of deposited code is established. \citet{trisovic2022} retrieved and executed
more than nine thousand R files from over two thousand Dataverse packages.
\citet{kranz} studies heteroskedasticity-robust inference using regressions recovered from deposits.
\citet{xuyang} evaluate an AI-assisted workflow for replication and reanalysis across many studies;
\citet{llmrep} demonstrate a replication prototype on one sociology paper.
These efforts recover or execute computations. Our principal screen joins command evidence to a
profile of the response without executing the deposited workflow; approximate refits are a separate
diagnostic. Detailed reanalyses, such as the treatment by \citet{beck} of temporal dependence in binary
outcomes, address aspects of an analysis that a marginal variable profile cannot recover.

\paragraph{Reporting checks and model choice}
Concerns about the reliability of published findings~\citep{ioannidis} have produced a family of
transparent screens: statcheck recomputes reported significance tests~\citep{statcheck}, GRIM tests
whether reported means are arithmetically possible~\citep{grim}, and Carlisle examines the
distributions in reported baseline tables~\citep{carlisle}; SPOT benchmarks automated detection of
listed manuscript errors~\citep{spot}. All of these read reported summaries. The profile here joins a
deposited variable's own empirical profile to the analysis that used it, which those screens do not
do. Choosing a model family is an established statistical task in its own right~\citep{gamlss}; our
screen does not choose, it records which combinations occur in existing analyses so that they can be
inspected. Its rubric draws on the event-count and fractional-response
literatures~\citep{king,papke,warton}, the case against log-transforming counts~\citep{ohara}, and
the debates around binary and ordinal responses~\citep{angrist,long}. Those literatures motivate the
use case; they do not make every selected combination an error.

\section{An application profile for analysis-scoped metadata}
\label{sec:profile}
The profile supports three operations: finding reported uses of a variable, joining those uses to
empirical evidence for a declared data scope, and selecting assertions by their provenance and review
state.

A column is not permanently ``the outcome'': the same variable can be the response of one analysis
and a predictor of another, so the metadata needs identities and relations, not a tag on a column.
Table~\ref{tab:profile} lays out the profile. Its core is the DDI-CDI 1.0 \texttt{InstanceVariable},
and the model also provides process and agent classes (\texttt{Activity}/\allowbreak\texttt{Step},
\texttt{Parameter}, human and machine agents) that the profile can reuse~\citep{ddicdi}. The role,
empirical-profile, and review predicates are application-specific and live in their own namespace; we
do not present them as approved DDI-CDI properties. PROV-O records generation and revision, and Web
Annotation points at the supporting locations in paper and code~\citep{provo,annotation}.
Nothing in the profile is specific to a repository platform, and in the pipeline only the harvest and
the pairing of deposits with papers read Dataverse fields. The inputs are delimited tabular files and
Stata or R scripts; the export is DDI-CDI, PROV-O, and Web Annotation graphs. Harvard Dataverse is
where the studied deposits live; any repository that exposes data files, code, and a link to the
published paper can substitute its own.

\begin{table*}[t]
\caption{The proposed application profile. ``Reported analysis'' denotes a specification described in
sources; it does not assert that an execution occurred. The example export implements a small fragment
of this profile, separate from the census pipeline.}
\label{tab:profile}
\small
\begin{tabular*}{\textwidth}{@{\extracolsep{\fill}}lp{0.70\textwidth}@{}}
\toprule
Object & Identity, scope, and content \\
\midrule
Dataset/file and variable & Dataset version, file identifier/checksum, and a stable DDI-CDI
\texttt{InstanceVariable} identity within that file; column name is an attribute, not the join key. \\
Empirical profile & Variable identity, data scope (which rows were profiled), statistics, inferred
class, missing-value policy, excluded values with their evidence, and profiler/rule version. \\
Reported analysis & Paper/table and code locations, response transformation and sample description;
estimator evidence is recorded separately from likelihood or link. \\
Analysis-role assertion & A claim that this analysis is headline, placebo, first stage, balance, or
robustness in this paper, supported by source evidence. \\
Variable-use assertion & Outcome or predictor in this analysis, linked to the actual variable;
exact-column identity and a derived-variable link remain distinct. \\
Extraction/review records & Evidence source, generator and configuration, version and time or declared
time uncertainty; separately attributed and scoped review decisions, retaining revision history. \\
\bottomrule
\end{tabular*}
\end{table*}

An empirical profile belongs to observations of a versioned variable. The full deposited column and a
reported estimation sample therefore have separate scope records; a restriction written in code is
not a verified row mask. A transformed response has its own identity and lineage through inputs and
a transformation step. For example, the logged battle-deaths column \texttt{lnbdb} can be linked to its raw count
\texttt{battledeadbest} without checking the logged response as though it were an untransformed count. A variable can also be an
outcome in a balance comparison and a predictor in another analysis. A paper-level summary of
headline outcomes can be materialized from these relations, with its derivation retained.

Three independent questions govern provenance. \emph{Evidence source} distinguishes paper text,
deposited code, codebook, data profile, and execution trace. \emph{Generation method} distinguishes
manual entry, deterministic extraction, and extraction by a large language model (LLM), including
the tool/model, prompt or rules,
configuration, and time. \emph{Review state} is unreviewed, accepted, rejected, or unresolved, with
reviewer, scope, and time on a separate decision record. LLM-generated claims remain LLM-generated
after review; parser output can also be wrong. Reviewing a metadata mapping is different from
endorsing a scientific specification. Conflicting and superseded assertions and decisions are retained;
queries select an explicit current view rather than silently replacing them with one authority.

\subsection{An example export and its queries}
The example is one testbed deposit, the hate-crime study of Section~\ref{sec:intro} (HRK5HI,
version 1.0; its analysis script is a separate deposit, VCYOJV, so the code evidence points there).
We export four of its reported analyses, located by their table in the source paper: the headline
regression (Table 2, column 1), the paper's own
quasi-Poisson robustness refit of the same outcome (Table 5, column 1), a placebo on hate crimes
against the majority (Table 4, column 1), and a pre-election balance comparison (Table B.2). The graph fixes the data file by its archive identifier and checksum, gives every column
a stable local identifier derived from that checksum, and attaches six empirical profiles computed
from the 280 deposited rows: the three responses, the derived difference \texttt{hc\_d}, and the two
difference-in-differences indicators. The analysis assertions come from the code-augmented fingerprint of Section~\ref{sec:textroute} and
keep that provenance; a separate set of review decisions,
attributed to an AI-assisted mapping review, accepts the mappings behind the headline alert and
leaves everything else unreviewed.

\begin{table}[t]
\caption{Analysis inventory returned from the example export. The robustness refit and the balance
comparison remain retrievable although neither produces an alert under the selected screen.}
\label{tab:inventory}
\small
\setlength{\tabcolsep}{2pt}
\begin{tabular*}{\columnwidth}{@{\extracolsep{\fill}}llll@{}}
\toprule
\shortstack[l]{location in\\source paper} & role & response & estimator \\
\midrule
Table 2(1) & headline & \texttt{hcam1} & linear \\
Table 5(1) & robustness & \texttt{hcam1} & quasi-Poisson \\
Table 4(1) & placebo & \texttt{hindu\_hc} & linear \\
Table B.2 & balance & \texttt{hvs} & t-test \\
\bottomrule
\end{tabular*}
\end{table}

Four queries read the graph, each a stricter view of the same relationships:
\begin{itemize}
\item \emph{Inventory}: every analysis with a role, an outcome use, and an estimator, alert or not.
Four rows (Table~\ref{tab:inventory}), including the robustness refit and the balance comparison,
which raise no alert.
\item \emph{All alerts}: analyses whose outcome use carries a model-choice alert, the profile's
record of a screening flag. In this example the rule flags a linear model on a count or share (the
clear tier) or on a binary or ordinal outcome (the debated tier); Section~\ref{sec:operator} defines
the rule and its tiers. Two rows: \texttt{hcam1}, profiled as a count, and the placebo response
\texttt{hindu\_hc}, a count with few distinct values that the classifier's heuristic reads as
ordinal (Section~\ref{sec:measured}).
\item \emph{Headline alerts}: the same, restricted to the headline role. One row, \texttt{hcam1}.
\item \emph{Reviewed}: headline alerts whose role, variable use, estimator, profile, and alert all
carry a current accepted review decision and no unresolved or rejected one. One row.
\end{itemize}

The headline query is shown in Figure~\ref{fig:query}, in the SPARQL query language for RDF
(\texttt{ap:} is the prototype's namespace; the shipped queries also exclude superseded assertions).
\begin{figure}[t]
{\footnotesize
\begin{verbatim}
SELECT ?v ?alert WHERE {
 ?r a ap:AnalysisRoleAssertion;
    ap:analysis ?a; ap:role ap:Headline.
 ?u a ap:VariableUseAssertion;
    ap:analysis ?a; ap:use ap:Outcome;
    ap:variable ?v.
 ?alert a ap:ModelChoiceAlert;
    ap:analysis ?a; ap:variableUse ?u.
}
\end{verbatim}
}
\caption{The headline query: every analysis with a headline role, its outcome use, and the alert on
that use, joined on the shared analysis}
\label{fig:query}
\end{figure}
Each pattern joins on the shared analysis \texttt{?a}: a role assertion saying it is the headline, a
variable-use assertion naming its outcome \texttt{?v}, and an alert on that use.

Three details show what the profile records and a flat table of flags would not. First, the headline
alert on \texttt{hcam1} sits next to the paper's own quasi-Poisson refit of the same column, recorded
as a robustness analysis that raises no alert. The census of Section~\ref{sec:atscale} records only
this deposit's two linear flags and nothing else (its script was read from the sibling deposit, as
Section~\ref{sec:atscale} states). A flag on its own would therefore drop the context
the paper itself supplies.
Second, the derived column \texttt{hc\_d} is exported as its own variable with a subtraction lineage
from \texttt{hcam1} and \texttt{hindu\_hc}, checked on all 280 rows, instead of being profiled as a
count; the audit wrapper of Section~\ref{sec:measured} leaves its class unresolved, because the profiler's
missing-value heuristic would otherwise strip the value $-1$ as a missing-value code. Third, review is scoped: accepting the headline mappings says nothing about the
placebo's, which is why the reviewed query returns one row where the alert query returns two.

A released script checks this one export in three ways.
\begin{itemize}
\item \emph{Interchange.} The script expands the JSON-LD, round-trips it through Turtle, and
compacts it again. The graph is the same at every stage: 477 triples, and every field of the input
document is found in the graph before and after the round trip. The four queries return the same
rows before and after.
\item \emph{Shapes.} The 26 triples that use DDI-CDI terms are validated on their own against the
DDI-CDI reference shapes. The application nodes are validated against their own shapes, which are
closed, so any property the profile does not declare is a violation.
\item \emph{Semantics.} Mutated copies of the graph must fail or change as the design says.
Validation fails when an assertion has no evidence, when an alert's profile belongs to a different
variable than its outcome, or when a role is stated on a column rather than an analysis. A
conflicting or superseded review stays in the graph, and the affected row drops out of the reviewed
query. Removing the alert nodes empties the alert queries and leaves the inventory unchanged.
\end{itemize}
These checks show that the declared relationships hold and that the graph survives interchange.
They do not measure extraction accuracy, which Sections~\ref{sec:atscale} and~\ref{sec:textroute}
do.

\section{Model-choice screening as a metadata query}
\label{sec:operator}

The screening use case joins two things a repository can recover about an empirical study: the
estimator used in a reported analysis and the observed characteristics of its response variable. We
map deposited commands to estimator families and join them to the inferred class of the response's values (the
\texttt{distributionKind} property of Section~\ref{sec:measured}); a rule anchored in the critique literature then selects combinations for
review. An estimator name alone says little about an analysis's distributional assumptions and
nothing about its inferential target. Ordinary least squares (OLS) does not require a Gaussian
marginal outcome, a linear
estimator can be read as a projection, and regression adjustment in experiments has a design-based
justification~\citep{lin2013}. A likelihood or link the authors state explicitly is separate,
evidence-backed metadata; finding \texttt{lm} or \texttt{regress} is not grounds for recording that
the authors assumed Gaussianity.

The screen selects linear estimators on variables classified as count, proportion, binary, or
ordinal. Its two tiers are routing categories, not verdicts: \emph{clear} covers counts and
proportions, where King and Papke--Wooldridge make the case, and \emph{debated} covers binary and
ordinal outcomes, where the linear model has prominent published defenders~\citep{king,papke,angrist,long}.
Deposited code supplies command-level evidence of a specification, not a trace of what ran.

\paragraph{What the two diagnostics establish}
An empirical profile summarizes the marginal values of a variable within one data scope. A
conditional model also depends on covariates, transformations, sample restrictions, and the
inferential target, none of which a column's values reveal: skewed integer ages can look like event
counts, and a bounded response that stays in the interior may admit a useful linear approximation.
Unlike the arithmetic checks of statcheck and GRIM~\citep{statcheck,grim}, this rule is a heuristic
screen, not a necessary condition for valid inference.

The refit diagnostic asks a narrower question: do reconstructed linear fits leave the response's
admissible range by more than numerical tolerance? Where they do, the fit cannot be read as an
in-domain conditional mean at those observations. That does not by itself disprove a linear
projection, a treatment-effect estimand, or a headline conclusion, and the census refits only
approximate the published estimation sample (Section~\ref{sec:censusrefits}).

\section{The measured side: empirical profiles linked to DDI-CDI}
\label{sec:measured}

\subsection{From value descriptions to empirical evidence}
A DDI-CDI description of a tabular file gives each variable a \texttt{SubstantiveValueDomain}
whose \texttt{recommendedDataType} comes from a controlled vocabulary aligned with XML Schema (XSD)
datatypes; the stock
generator emits the common terms, Integer, Double, Boolean, Date, and String. The vocabulary also has
integer subtypes such as \texttt{NonNegativeInteger} and \texttt{PositiveInteger}, and DDI maintains
a separate \texttt{NumericType} vocabulary with an explicit \texttt{Count} term, so the standard is
not silent about counts. The datatype is nevertheless the wrong cut for the screen: a count and a
1-to-5 Likert item are both non-negative integers, and a bounded proportion and a GDP column are both
\texttt{Double}. The observed support and shape of the values separate these cases.
Table~\ref{tab:smoke} shows the stock generator and the extension side by side on a five-column
verification file.

\begin{table*}[t]
\caption{Stock datatype, the DDI-CDI measurement scale (\texttt{classificationLevel}), and the extension on a
five-column verification file. XSD datatypes collapse count/binary/ordinal into Integer and
proportion/heavy-tailed into Double; the standard's own scale level separates those, but still maps count,
proportion, and heavy-tailed all to Ratio; \texttt{distributionKind} restores the distinction the screen
needs.}
\label{tab:smoke}
\small
\begin{tabular*}{\textwidth}{@{\extracolsep{\fill}}llll@{}}
\toprule
variable & datatype & \texttt{classificationLevel} & \texttt{distributionKind} \\
\midrule
battle\_deaths & Integer & Ratio & count \\
treated & Integer & Nominal & binary \\
likert\_1to5 & Integer & Ordinal & ordinal \\
vote\_share & Double & Ratio & proportion \\
gdp\_pc & Double & Ratio & heavy-tailed \\
\bottomrule
\end{tabular*}
\end{table*}

\paragraph{Scale is not distribution.} DDI-CDI does carry more than the XSD datatype. A
\texttt{ValueAndConceptDescription} records a \texttt{classificationLevel} drawn from the model's
\texttt{CategoryRelationCode} enumeration (nominal, ordinal, interval, ratio, continuous), a
measurement scale that descends from the \texttt{@nature} attribute of DDI-Codebook. We reuse it
rather than reinvent it: our nominal, ordinal, and continuous distinctions \emph{are}
\texttt{classificationLevel}, and the extended generator emits it next to \texttt{distributionKind}.
Scale is still too coarse for the screen. A count and a bounded proportion are both ratio-scale, so
\texttt{classificationLevel} gives them the same value, yet their supports call for different review
questions and different alternatives, count models such as Poisson or negative binomial for the one
and fractional-response models for the other. The operational class that sits \emph{below} the scale
level, count against proportion against heavy-tailed, is recorded by no property the generator
emits; it is what \texttt{distributionKind} adds. We keep the standard scale level and add the
observed evidence the screen uses.

\subsection{The extension}
We fork the \texttt{rdm-integration} toolchain's streaming DDI-CDI generator \citep{rdmint}
(\texttt{cdi\allowbreak\_generator\allowbreak\_ext.py}, from
\texttt{cdi\allowbreak\_generator\allowbreak\_jsonld.py}). The stock generator
already streams each file once, row by row, maintaining per-column state: candidate-type flags,
approximate distinct counts via HyperLogLog, and inference of each column's structural kind (identifier, measure,
dimension, attribute). The extension adds to the same pass the numeric evidence a distribution class needs
(minimum and maximum, counts of negative and unit-interval values, running first three moments for
mean, variance, and signed skew, and a capped set of small values) and maps it to a class with
transparent rules.

The census classifier, the decision sequence shipped in the extension, is as follows. Non-numeric
columns are categorical. For integer columns whose captured value set is small (at most 25 values)
the classifier is aware of sentinels, that is, missing-value codes, a step added after the first
pilot (Section~\ref{sec:atscale}).
It first strips a fixed list of survey missing-value codes (98, 99, 999, 9998, 9999, and the negative
codes $-1$, $-7$, $-8$, $-9$, $-99$, $-998$, $-999$, $-9998$, $-9999$), along with any value outside
$[-90, 90]$, and reads the remaining core. There, at most two distinct values gives binary. An
integer scale with minimum $\ge 0$, maximum $\le 12$, and at most 12 distinct values gives ordinal,
so a Likert or rating scale is recovered even when missing-value codes inflate the raw range.
Otherwise the gates run in order:
\begin{itemize}
\item at most two distinct values: binary;
\item non-negative integers with maximum $\le 12$ and at most 12 distinct values: ordinal;
\item non-negative integers with signed skew $\ge 1.0$, more than 4 distinct values, and maximum
below $10^{6}$: count;
\item all values in $[0,1]$ with more than 12 distinct values: proportion, whereas a few-valued
unit-interval column (at most 12 distinct values) is ordinal, the signature of a Likert or rating
item min-max rescaled into $[0,1]$ rather than a genuine bounded share;
\item non-negative values with skew $\ge 2$: heavy-tailed; otherwise continuous.
\end{itemize}
These gates are heuristics. Skew and magnitude do not uniquely separate counts from ages or amounts,
and a small unit-interval support does not separate a rescaled rating from a share with a small
denominator; the thresholds route review candidates, they do not settle those semantics. The class
is emitted as a \texttt{distributionKind} property on each variable's
\texttt{Substantive\allowbreak Value\allowbreak Domain} in the generated JSON-LD, alongside the
standard datatype and never replacing it. That is the generator's export location; because the class
can change with the observations while the value domain does not, the application profile attaches
it to a separate empirical-profile entity with a data scope (Section~\ref{sec:profile}).
Deterministic here means repeatable rules, not certain semantics: the categories mix support,
empirical shape, and inferred substantive type, and they are not fitted probability distributions.
Computation needs one pass, bounded per-column state, and no model calls. The gates reflect
development on pilot adjudications (Section~\ref{sec:atscale}).

The extension reuses the generator's existing file pass, so a repository that adopts it produces this
evidence during metadata generation and can reuse the profiles across analyses.

\subsection{Three characterized failure modes}
\label{sec:failures}
Marginal, single-column profiling has three failure modes we can name precisely, all met in real
data. \textbf{Scale-masked proportions}: a bounded rate stored on a 0--100 scale (an
electrification rate in the testbed) profiles as continuous, so an OLS-on-proportion mismatch is
missed, a false negative. \textbf{Symmetric small counts}: a low-range, low-skew integer count is
indistinguishable from a rating scale on the marginal alone; species-richness counts (2--11, skew
1.18) profile as ordinal, so the Gaussian ANOVA is still flagged (ordinal $\to$ linear also triggers
the rule) but lands in the debated tier instead of the clear count tier.
\textbf{Rescaled Likerts read as proportions}: a single survey item min-max rescaled into $[0,1]$ is
bounded like a share but is substantively an ordinal rating; the revised gate routes few-valued
unit-interval columns to the debated ordinal tier, which removes this source of clear-tier flags at
the price of routing small-denominator shares there too (Section~\ref{sec:atscale}). All three modes
follow from reading one marginal at a time, and all three are visible in the metadata itself (a
maximum of 100, a symmetric small support, a handful of evenly spaced values in $[0,1]$). Codebook
units and category labels could resolve them; we leave that as future work rather than patch the
taxonomy ad hoc.

\paragraph{A sentinel counterexample and regression checks}
The sentinel step has a concrete counterexample: a legitimate count column with values
$\{0,50,100\}$ loses the 100 to the sentinel strip and comes back as binary. It is a synthetic case,
not a measured failure rate. The released regression checks cover small and large counts, two
observed count values, valid 99 and 100 counts, genuine versus declared sentinel negatives,
percentages, small-denominator shares, rescaled ratings, constant columns, and all-missing columns;
they document the ambiguities as well as this failure.

For the application-profile example, \path{profile_audit.py} wraps the census classifier,
unchanged, and keeps the original support (or an explicit marker when it exceeds 25 distinct
values), the statistics, and the values the census classifier's sentinel heuristic would have excluded. It drops
observations only when a codebook or DDI declaration marks them as sentinels, with that evidence
attached. A class the census classifier reached only by stripping a listed code or an out-of-range
value is marked unresolved unless such a declaration names that value. Constant and all-missing
columns are unresolved as well. The wrapper makes exclusions reviewable and reversible; it leaves
the census results untouched and claims no better classification accuracy.

\section{A benchmark for metadata recovery}
\label{sec:testbed}

\begin{table*}[t]
\caption{The testbed's norm map: recorded outcome class (rows) by deposited estimator family
(columns), counts of cases. The mismatch strata were selected to occupy the linear column; the
table describes benchmark composition rather than the prevalence of model choices.}
\label{tab:norm}
\small
\begin{tabular*}{\textwidth}{@{\extracolsep{\fill}}lcccc@{}}
\toprule
outcome $\backslash$ estimator & linear & count & binary & proportion \\
\midrule
count        & 8  & \textbf{6} &   &   \\
proportion   & 12 &   &   & \textbf{1} \\
binary       & 6  &   & \textbf{1} &   \\
ordinal      & 6  &   &   &   \\
\bottomrule
\end{tabular*}
\end{table*}

\begin{table*}[t]
\caption{Ten of the 32 benchmark mismatch cases (full 40-case testbed released with DOIs, evidence notes,
and measured classes). Tier (a routing category, Section~\ref{sec:operator}): clear = the King/Papke--Wooldridge/Warton--Hui
strata; debated = the LPM and ordinal-as-continuous strata. A listing records benchmark membership,
not an assessment of the study's conclusions.}
\label{tab:mismatch}
\footnotesize
\setlength{\tabcolsep}{1.5pt}
\begin{tabular*}{\textwidth}{@{\extracolsep{\fill}}lllll@{}}
\toprule
field & outcome & measured & estimator & tier \\
\midrule
pol.\ sci.\ (India) & hate-crime incidents & count & \texttt{lm.cluster} & clear \\
public policy (US) & firearm background checks & count & \texttt{regress} & clear \\
int.\ relations / civil war & battle deaths & count & log-OLS (\texttt{reg}) & clear \\
environmental policy & utility disconnections & count & \texttt{reg} & clear \\
pol.\ sci.\ (S.\ Africa) & ANC (African National Congress) vote share & proportion & \texttt{reg} & clear \\
pol.\ sci.\ (Colombia) & municipal turnout rate & proportion & \texttt{xtreg} & clear \\
dev.\ econ.\ (randomized trial) & sisters-in-school share & proportion & \texttt{reg} & clear \\
pol.\ sci.\ (US) & individual turnout (0/1) & binary & \texttt{lm.cluster} & debated \\
pol.\ sci.\ (survey exp.) & 4-cat.\ agreement scale & ordinal & \texttt{reg} & debated \\
ecology & species richness per plot & count (prof.\ ord.) & \texttt{aov} & debated$^{\ast}$ \\
\bottomrule
\end{tabular*}
\par\smallskip
{\footnotesize $^{\ast}$Recorded as a count outcome, the clear tier under the rubric (King, O'Hara--Kotze); the marginal profiler reads the
low-skew 2--11 support as ordinal, so the flag is raised at the debated tier
(Section~\ref{sec:failures}).}
\end{table*}

\subsection{Construction and verification}
The testbed supplies reference estimator--outcome links and operational screening labels for the
acquisition experiments. To our knowledge, and after a deliberate search during its construction, no
labeled corpus of published analyses existed for this class of estimator--outcome combinations: per-paper model-family critique
is a methodologists' genre rather than a routine check, and King counted the studies without naming
them. The testbed (\texttt{operator2\_testbed\_v2.csv}, released) contains 40 cases drawn from 30 public
Harvard Dataverse replication packages, each a (dataset, outcome variable, estimator) triple with a
recorded outcome class, a benchmark label, and a note recording the evidence. The label marks a
pattern from the critique literature; it is neither the census rule's selection nor an adjudicated
verdict, three definitions of a positive that Section~\ref{sec:atscale} keeps apart. Candidates
were located through the Dataverse Search API along five deliberate strata (count outcomes under
linear models; count outcomes under count models; proportion and rate outcomes; binary and
ordinal outcomes; cross-domain breadth beyond political science), with inclusion requiring a public,
readable tabular file and deposited analysis code naming the estimator and the outcome. Membership
therefore reflects deposit completeness, not an assessment of the study. It holds
32 mismatches (8 count, 12 proportion, 6 binary, 6 ordinal, all under linear estimators)
and 8 matched-family controls (Poisson on media counts, negative binomial and zero-inflated negative
binomial on terrorism and dispute counts, beta regression on coded discourse shares, logit on a binary
vignette outcome). Fields span political science, international relations, conflict studies, public
policy, public health, development economics, energy and environmental policy, history, education, and
ecology; the set cannot be reduced to one literature. Table~\ref{tab:norm} summarizes the strata;
Table~\ref{tab:mismatch} shows ten of the mismatch cases.

In July 2026 every case was re-checked against the live archive. Each DOI resolves with matching
metadata. Each outcome column re-profiles to its recorded class, with one disclosed exception
(Section~\ref{sec:tie}); where no deposited column stores the outcome, we built it the way the
deposited script builds it. Each estimator--outcome link was confirmed by reading the deposited
Stata do-files and R scripts. Labels are anchored to the published critique literature (King for
counts~\citep{king}; Papke--Wooldridge for proportions~\citep{papke}; Warton--Hui for ecological
proportions~\citep{warton}; Long and Angrist--Pischke framing the debated binary/ordinal
tiers~\citep{long,angrist}). They are operational benchmark labels, not independent scientific
adjudications of each paper. The check caught one labeling error of ours: an African
legislative-bills dataset filed as a count-under-OLS exemplar in fact deposits \texttt{nbreg} code, a
count model, and was moved out of the mismatch set; the withdrawn case stays in the audit trail.

\subsection{Checks on the recorded labels}
\label{sec:tie}
Because of how the benchmark was built (Table~\ref{tab:norm}), the detector and the bare rule,
\emph{flag a linear estimator on a non-continuous outcome}, are equivalent on it. With the recorded
outcome classes both return 32 positives and eight negatives, precision and recall 1.0. That checks
the implementation against the labels; it says nothing about whether flagged analyses are in fact
misspecified. What the metadata adds over the bare rule is the tier routing and the linked evidence
for each candidate.

The raw testbed profiler (\path{profile_outcome.py}), which emits the same label set under its
own earlier thresholds, agrees with the recorded class on 39 of 40 cases. The exception is the
scale-masked electrification rate, recorded at its verified proportion class; run from the profiler's raw output instead, the pipeline misses that case and recall drops
to $31/32$.
The census next tests whether those reference links can be recovered from uncurated deposits.

\begin{table*}[t]
\caption{Illustrative re-analyses. Coefficients are on each model's own scale: additive outcome units
for OLS, log expected counts for Poisson/negative binomial, and log odds of the conditional mean for
fractional logit. Log-OLS models the logged response. Magnitudes across columns are not comparable.
The final column reports the specific observed domain issue, not inferential validity.}

\label{tab:reanalysis}
\small
\begin{tabular*}{\textwidth}{@{\extracolsep{\fill}}llll@{}}
\toprule
case & linear reconstruction & alternative specification & domain diagnostic \\
\midrule
hate crimes & OLS $+1.78$, $p{=}0.001$ & Poisson $+1.96$, $p{=}0.004$ & 29\% $<0$ \\
battle deaths & log-OLS, sig. & neg.\ binomial, sig. & none (no zeros; min 900) \\
ANC vote share & $+0.21$, $p{<}0.001$ & frac.\ logit $+1.07$, $p{<}0.001$ & 0\% (in range) \\
\bottomrule
\end{tabular*}
\end{table*}

\subsection{Illustrative refits and analysis context}
\label{sec:reanalysis}
For three clear-tier benchmark cases we fitted alternative specifications to the deposited data
(Table~\ref{tab:reanalysis}; \texttt{reanalysis.py}, statsmodels). None reverses the published
significance. Coefficients from OLS, count models, and fractional logit live on different scales, so
their magnitudes say nothing about a stronger or weaker effect under the alternative family, and a
marginal class does not single out one alternative as the right one.

The reconstructed hate-crime OLS fit has negative fitted counts for 29\% of its observations, a
domain violation of the conditional mean in the headline specification; the published paper also
reports quasi-Poisson robustness checks~\citep{basu}. That headline is a difference-in-differences
with state and year fixed effects, where a linear estimator is standard, so the negative fitted
values concern the conditional mean at those observations and not the treatment-effect estimand. The
battle-deaths analysis models \texttt{lnbdb}, the log of \texttt{battledeadbest}~\citep{lacina};
the deposited counts have no zeros (minimum 900), and linking the two variables neither turns the log
response into a raw count nor shows a domain violation. The ANC vote-share reconstruction keeps its
fitted values within $[0,1]$~\citep{dekadt}. The latter two are flags without the diagnostic, not
failures of the published inference.

\begin{figure*}[t]
\centering
\includegraphics{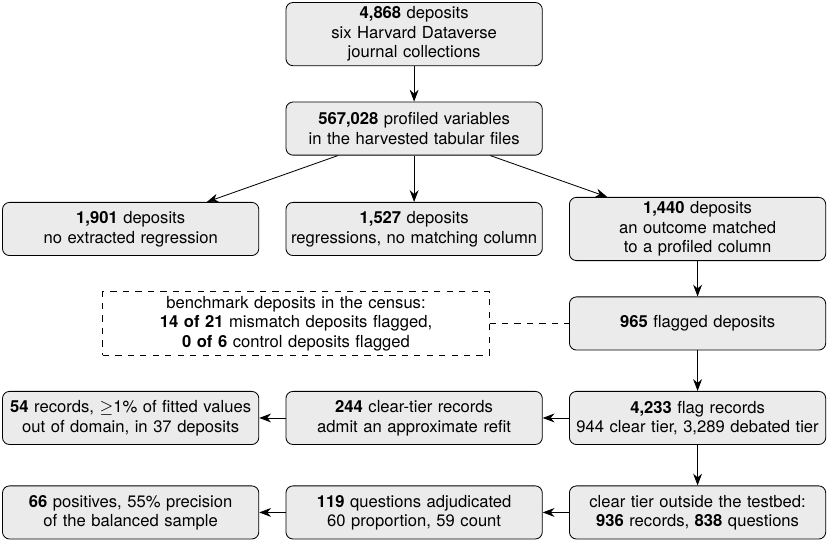}
\caption{The census as a flow, with the counts the released driver regenerates: harvested deposits and
profiled variables, the split by extracted regressions and matched outcome columns, flagged deposits
with the benchmark overlay, flag records by tier, the refit branch (records admitting an approximate
refit, and those with at least 1\% of fitted values out of the response domain), and the clear-tier
pool outside the testbed with the adjudicated sample and its positives}
\label{fig:census}
\end{figure*}

\section{Metadata coverage and screening at repository scale}
\label{sec:atscale}

The largest experiment measures what the profiler and code extractor recover when the
screen runs over six journal collections. The archive was processed without special treatment
for benchmark deposits, with one exception: the hate-crime deposit of Section~\ref{sec:profile} ships
no code of its own, and its census copy included the analysis script deposited alongside it in VCYOJV. The classifier's gates were developed on pilots drawn from the same corpus,
so the precision reported below is a same-corpus, post-development estimate; the development record
follows the results.

\paragraph{What is measured}
The census pipeline (\texttt{atscale\_pipeline.py}) joins the inferred
\texttt{distributionKind} to estimator evidence extracted from deposited \texttt{.do} and
\texttt{.R} files, then applies the rule of Section~\ref{sec:operator}. What it measures is the
coverage of that join, benchmark recovery, candidate volume, and precision on a sample.

\paragraph{Corpus}
The run covers \textbf{all 4{,}868 datasets} that six Harvard Dataverse journal collections held when
harvested in the first week of July 2026:
American Journal of Political Science (AJPS), Journal of Politics (JOP), American Political Science
Review (APSR), Political Analysis, International Studies Quarterly (ISQ), and Comparative Political
Studies (CPS). The pool directory also held fourteen bare-tag directories left by the testbed preparation,
ten of them duplicates of harvested deposits and none yielding a regression; they are excluded from
every count. Harvesting used the Dataverse native API with no tabular-file or code-presence
precondition, yielding \textbf{567{,}028 profiled variables}. The corpus includes 36 of the 40
testbed cases. Counted by deposit (one Dataverse dataset), it includes 21 of the 24 deposits that
carry a mismatch case and six of the seven that carry a control case; one deposit carries both. Coverage and flag counts describe this harvested population;
precision is estimated from a sample of its candidates.

\paragraph{Pipeline}
For every dataset: harvest the tabular files and deposited code via the Dataverse native API (the only
repository-specific step); generate
enriched DDI-CDI with per-variable \texttt{distributionKind} (\path{cdi_generator_ext.py}); extract
(estimator, outcome) pairs from the deposited Stata and R scripts by the keyword mapping, including
handling of recognized \texttt{glm} family arguments to distinguish Poisson or binomial generalized
linear models (GLMs) from the linear family; match each extracted outcome name against the profiled
columns; flag every linear estimator
whose outcome profiles as count, proportion, binary, or ordinal, graded clear (count, proportion) or
debated (binary, ordinal). The harvest has fixed limits: it takes tab and csv data files and do, R,
and Rmd scripts, skips any file over 25~MB, and keeps at most eight files of each kind per deposit in
the archive's listing order, so other formats, larger files, and later-listed scripts are never seen.
The keyword mapping is a fixed pattern over estimator command names with a short list of accepted
Stata prefixes, so a command form outside that list, such as the survey prefix \texttt{svy:}, is not
extracted. On the 36 testbed cases present in the census, the pipeline locates the reference outcome
variable in the profiled data for \textbf{30 (83\%)}; the six misses are outcomes that exist in no
profiled column, because the analysis code builds or renames them or because their data file fell
outside the harvest limits. Figure~\ref{fig:census} draws the flow from the harvested deposits to the
adjudicated sample.

\paragraph{Benchmark recovery}
Recovery is reported per deposit, with case-level outcome recovery alongside. Processed with the
rest of the census, the pipeline flags \textbf{14 of the 21 in-census mismatch datasets} (67\%; 14 of
the full 24) and leaves \textbf{all 6 in-census control-bearing deposits unflagged}. A control
``pass'' here means no flag anywhere in the deposit, not a judgment on its named control analysis;
one deposit carries both a control and a mismatch, so its absence of flags is at
once a control pass and a missed mismatch, and six deposits do not establish specificity. At case
level, 21 of the 29 in-census mismatch cases sit in a flagged dataset and 19 have the exact recorded
outcome flagged. The seven missed datasets are fully accounted for. Three are harvest limits. In
one deposit the analysis script is the 29th of 37 code files and falls outside the eight-file cap. In
one the data files exceed the size cap, so no variable was profiled. In one, the harvested copy of the
script yielded no regressions, and its data file also exceeds the size cap, so the outcome could not
have been matched in any case. Re-fetching that deposit without the cap and running the same screen
profiles its 1.26~GB voter-level file and flags the recorded outcome, a binary turnout indicator
under \texttt{lm}. The miss is the cap, not the screen. One is the keyword mapping: every regression in the testbed case's script
carries the \texttt{svy:} prefix. Two build the outcome in code, a dichotomized attitude scale and a
recoded vignette item, so neither outcome is a stored column. One is the scale-masked electrification rate of
Section~\ref{sec:failures}. Six of the seven are acquisition and matching limits of this pipeline,
not the screen.

\paragraph{Coverage and flag volume}
\textbf{1{,}901 of the 4{,}868 datasets
(39\%) yield zero extractable regressions} under these harvest rules and this mapping and are
unflaggable by construction; a further 1{,}527
(31\%) yield regressions whose outcomes match no profiled column; the flaggable base, datasets with at
least one matched outcome, is \textbf{1{,}440 (30\%)}. Flags land on \textbf{965 datasets}: \textbf{198
per 1{,}000 deposits} over all 4{,}868, and \textbf{67\% of the flaggable base}. The first rate
describes the yield across all deposits; the second conditions on the pipeline recovering at least
one outcome link.
In total \textbf{4{,}233 flags}: \textbf{944 clear-tier} (count and proportion, the King and
Papke--Wooldridge cases) and 3{,}289 debated.

\paragraph{Precision on sampled questions}
The clear tier outside the testbed can be counted three ways, and the figures below use all three.
It holds \textbf{936 flag records} (758 proportion, 178 count); a record is one estimator on one
outcome, so a script that fits \texttt{lm}, \texttt{lm\_robust}, and \texttt{lmer} on the same column
contributes three. Collapsed to one question per (dataset, outcome) pair, it holds 838 questions
(671 proportion, 167 count); the question is the adjudication unit. From the records, 60 proportion
and 60 count were drawn at random and collapsed to questions, which removed one duplicate:
\textbf{119 questions}, 60 proportion and 59 count. Two AI-assisted sessions adjudicated them
against the deposited data and code, following the evidence protocol of Section~\ref{sec:testbed}.
A positive requires that the code uses the identified variable as an outcome and that its
substantive reading as a count or share meets the rubric; the rubric also rejects transformed
responses and specified cases where a linear approximation is defensible. The metric therefore
judges screen candidates, not extraction alone. Precision is \textbf{66 of 119, 55\%} (Wilson 95\%
CI $[0.47, 0.64]$): \textbf{proportions 37 of 60, 62\%} (CI $[0.49, 0.73]$) and \textbf{counts 29
of 59, 49\%} (CI $[0.37, 0.62]$).
Read from the archive up rather than from the metric down, the per-type figures suggest that of the
838 clear-tier questions the rule raises outside the testbed across six journals' replication
archives, roughly half survive a rubric
that also discards transformed responses and defensible linear approximations. That is several
hundred deposited analyses in which a linear estimator meets an untransformed count or share outcome,
each a review candidate with its data and code attached, not a verdict.

Three definitions of a positive are in play. Benchmark membership follows the critique literature
and includes one log-OLS fit on a count (Table~\ref{tab:mismatch}). The census rule selects a linear
family on a column whose profile is a count or share, matched by name; a logged response enters only
when its name is the raw column's name with an \texttt{ln} or \texttt{l\_} prefix, and the flag then
names the raw column. The adjudication rubric credits a flag only when the modeled response is the
identified column, untransformed, so such a flag counts against precision. Recovery thus asks whether
the rule reaches benchmark deposits, precision whether its candidates survive the rubric; the
logged-count pattern is in the benchmark but not among the rubric's positives, and needs a
transformation-aware check of its own.

The 55\% is precision on a sample balanced by type. The pool is not balanced, four proportion
questions to one count question, so the per-type figures are the ones that speak to the
archive. The record counts cannot serve as weights, because the sample is counted in questions;
weighting the per-type figures by the question counts instead gives about 59\%, and we report the
balanced figure. The draw was made at record level, so a question with several records had a higher
chance of selection; 15 of the 119 sampled questions carry two or three. Weighting each sampled
question by the inverse of its record count gives 63\% and 50\%, so the 59\% is an approximation,
not an inclusion-weighted estimate. The count false positives that remain are right-skewed integer covariates
(ages, membership tenures), list positions and ranks, and large near-continuous tallies judged
defensible under a linear approximation.

\paragraph{Development corrections}
Rule development and evaluation used the same corpus. The first adjudication pass on an initial
486-dataset pilot returned 11/50 positives (22\%). Its false positives included survey
missing-value codes, feeling thermometers, ages, and monetary amounts classified as event counts.
This prompted sentinel handling and the skew, distinct-count, and magnitude gates. A later overlap
pilot showed near-chance rater agreement, with disagreements concentrated on single survey items
rescaled into $[0,1]$: treating these as bounded proportions placed them in the clear tier, while
treating them as ordinal ratings placed them in the debated tier. The few-valued unit-interval gate
was added before generating the final census pool, from which the 120-record adjudication sample
was drawn. The hardened rule had scored 64\% overall and 69\% on proportions on the 486-dataset
pilot; that sample was almost all proportions, the better stratum, and the gate changed between the
two pools, so the pilot and census figures differ in both mix and rule and do not isolate a change
in performance. Two of the census sample's 119 unique questions also occur in the released earlier pilot verdicts. The superseded
overlap verdict files were not retained, so the record establishes neither a disjoint held-out
evaluation nor an independently timestamped rule freeze. These are documented development
decisions; the surviving record cannot show whether the changes improved generalization.

\paragraph{Adjudication reliability}
Two AI-assisted sessions (Anthropic's Claude), directed by the pipeline's author under a shared
written rubric, inspected deposited data and code. The primary session adjudicated all 120 sampled
records; the second, blind to those verdicts, adjudicated 40. On the 39 unique overlap questions the
two sessions agree on 37 (94.9\%), Cohen's kappa 0.90. One disagreement is whether a state-level
female-candidate proportion is a boundary share or a defensible interior index. The other is the
one duplicated outcome, sampled under two
estimators and judged inconsistently by the second session; counting that self-contradiction as a
disagreement gives 0.90, collapsing the duplicate favorably would give 0.95, and the released
calculation keeps the former.

Removing the duplicate leaves 119 unique questions, and the pool builder now deduplicates (dataset,
outcome) before sampling. The second session's 21 of 40 positives (52.5\% on raw records) is close to
the primary estimate. Agreement measures the consistency of two AI-assisted applications of one
rubric; an error both share would not show up as disagreement. All 160 verdict records and their
evidence notes are released.

\subsection{Approximate refits of census flags}
\label{sec:censusrefits}
For each clear-tier flag the archived implementation (\texttt{refit\_broken.py}) attempts linear
refits using recognizable deposited terms. It accepts plain Stata \texttt{reg}/\texttt{regress} and
selected R linear fitting calls, plain covariates, factor terms, and numeric R interactions. It skips
fixed-effects commands, instrumental-variable and mixed estimators, Stata \texttt{if}/\texttt{in}
and bracketed weights, and R \texttt{subset} arguments. Factor dummies in otherwise plain OLS remain
eligible. The data file is chosen as the largest CSV/tabular file containing the required column
names; the fit uses complete cases from at most 300{,}000 rows, requires at least 30 rows and three
times as many rows as parameters, and caps factor expansion at 300 levels.

These are \emph{approximate reconstructions}, not reproductions of the deposited workflow. The data
file is not resolved from the program's own loading statements; earlier transformations and
filtering, the original missing-value conventions, R weights and contrasts, and some Stata options
and grouped execution are not reconstructed; an intercept is inserted. The original estimation sample
is therefore not certified. The archived output keeps each selected fit's row and parameter counts,
its violating share, and the number of successful refits, but not every formula, source location,
file checksum, or row mask; an execution-based result would need that fuller record.

The boundary tolerance is $\epsilon=10^{-6}$ in response units: a count violates below $-\epsilon$,
a proportion below $-\epsilon$ or above $1+\epsilon$, and the observed response itself must lie
within that tolerance of the stated domain before any refit counts. The implementation tries at most
eight matching command records per flag and stores the largest violating share among the successful
refits, so the reporting unit is a flagged (dataset, estimator, outcome) record, not a unique
specification. Of 944 clear-tier records, 244 (26\%) admit at least one refit, 718 successful refits
in all. \textbf{54 of those 244 records (22\%), across 37 datasets, have a refit with out-of-domain
fitted values for at least 1\% of its estimation observations}: 17 of 52 count records (33\%) and 37
of 192 proportion records (19\%). Among these 54 the median violating share is 6\% and the maximum
35\%; counting any violation at all selects 80 of 244 (33\%). The hate-crime flag is that maximum, the worst of its six
reconstructed specifications; its own re-analysis script reproduces the 29\% of the single
specification in Table~\ref{tab:reanalysis}. These are diagnostics of reconstructed fits; given the coverage and the approximations
above, they are not a count of disproved published analyses.

\section{Recovering analysis evidence from papers and code}
\label{sec:textroute}

Reported-analysis evidence becomes metadata through the analysis fingerprint: the linked paper is
distilled, in one model call, into fixed-format records; the prompt requests one per reported
analysis, and completeness is not measured. Each record carries
the outcome verbatim, the estimator family as the paper states it (or ``not stated''), an analysis
role (headline, placebo, first stage, balance, robustness), and one column picked from the deposit's
variable-level metadata as the outcome. The prompt requires that the picked column's type and range
be consistent with the outcome, and it requires ``not in the metadata'' as an available answer. This section
measures that layer on the verified testbed.

Pairing each deposit with its published paper uses only repository metadata. Of the testbed's 30 deposits, 28 resolve to their
published papers, 8 through the DOI in the Related Publication field and the rest through the
deposit-title convention (``Replication Data for: $\langle$title$\rangle$'', stripped and matched by
containment against Crossref); the title rule leaves two deposits unresolved, one unpublished and one
whose paper we recovered by hand. Full text was obtainable for 28 papers. The unpublished deposit and
one study published as a book have none, which leaves 37 of the 40 testbed rows scorable. Two of
the 37 have outcomes that no deposited column carries. Both belong to one deposit, a vignette
experiment whose data file is the raw survey export with one column per vignette item; the deposited
script builds the two outcomes at run time, a 0/1 acceptance from the item responses and a four-point
agreement scale recoded from six argument items. Nothing is missing from the deposit. Reading the script recovers both rows, which is how they
entered the testbed; the census keyword join does not (Section~\ref{sec:atscale}), and the column
link alone has nothing to point at. The prompt requires ``not in
the metadata'' as an option, the models return it for these two, and the scorer counts that as no
link, so 35 outcome links are recoverable.

\begin{table}[t]
\caption{The analysis fingerprint on the 37 text-bearing testbed cases: one call per pair, identical
prompt, single run. Exact outcome = the recorded column, picked as an analysis outcome; derived outcome = an accepted
twin of it; outcome + family = an outcome link plus the code's estimator family. Code-added rows add the deposit's scripts to the same
call. Comparators: the earlier two-step pilot, and the code layer the labels were read from, which
scores 37 by construction and exists only where parseable code is deposited.}
\label{tab:textroute}
\small
\setlength{\tabcolsep}{2pt}
\begin{tabular*}{\columnwidth}{@{\extracolsep{\fill}}lccc@{}}
\toprule
model & \multicolumn{1}{c}{\shortstack[c]{exact\\outcome}} & \multicolumn{1}{c}{\shortstack[c]{derived\\outcome}} & \multicolumn{1}{c@{}}{\shortstack[c]{outcome\\+ family}} \\
\midrule
\multicolumn{4}{@{}l}{\emph{text + metadata, no code}} \\
claude-haiku-4-5 & 25 & 2 & 23 \\
claude-sonnet-5 & 33 & 1 & 27 \\
claude-opus-5 & 34 & 1 & 28 \\
claude-fable-5 & 33 & 1 & 26 \\
\multicolumn{4}{@{}l}{\emph{text + metadata + code}} \\
claude-haiku-4-5 & 26 & 0 & 25 \\
claude-sonnet-5 & 33 & 1 & 34 \\
claude-opus-5 & 34 & 1 & 34 \\
claude-fable-5 & 34 & 1 & 34 \\
\multicolumn{4}{@{}l}{\emph{comparators}} \\
two-step pilot (haiku) & \multicolumn{2}{c}{14 exact+derived} & 11 \\
code layer (label source) & -- & -- & 37 \\
\bottomrule
\end{tabular*}
\end{table}

We ran the identical prompt through four models on the paper's full text plus the deposit's
variable-level metadata (per column its name, its label where a codebook shipped, type, range,
distinct and non-missing counts), never code or data. For one deposit the metadata was rebuilt
from a re-fetch without the harvest size cap: its 1.26~GB voter-level file, which the repository
never ingested as tabular, holds the recorded outcome column and was profiled in full. We then ran the same four models with the
deposit's analysis scripts added as a third input, the code-augmented arm. Because the testbed's
estimator labels were derived from that code, the code-augmented arm measures extraction fidelity; it
tests enrichment where code is available, and the text-only arm tests the same task without it.
Each row is scored on two sides. On the outcome side, a row is \emph{exact} when the column picked as
the outcome of a reported analysis is the testbed's recorded column and \emph{derived} when it is an
accepted twin of that column, a recoded or transformed version of the same variable, such as the
logged battle-deaths column for its raw count; exact plus derived is the outcome link. A pick made only
as a treatment or predictor does not count, and a match verifies column and family, not the table
or specification the label tracks. On the estimator side, a row is recovered when the
stated family equals the deposited code's family. \emph{Outcome + family} counts rows with an outcome link
and a matching family, so it can never exceed the outcome link. An alias or derived link is not
evidence that the raw column and the modeled response share a support or a scale.
Table~\ref{tab:textroute} reports the grid.

The pattern is simple: the three larger models identify the outcome column almost equally well
with or without code, and code mainly raises estimator recovery. On the outcome side, Opus links
all 35 linkable rows in both arms (34 exact, one derived); Fable 34
from text and 35 with code; Sonnet 34 in both; Haiku 27 from text and 26 with code. Haiku names
three more of the recorded columns, and Fable one, only in a treatment or predictor record, which
the grid does not credit. Haiku's two arms do not link the same rows: with code, one of its two
derived matches becomes exact, the other is lost, three exact links are lost and three others are
gained. On the estimator side, text and metadata give the three
larger models 26 to 28 of 37. Most rows they miss are analyses whose family the paper never states:
balance tables, eligibility and first-stage regressions, a placebo table that inherits the headline's
estimator, and survey-weighted comparisons. For these the models write ``not stated''. The rest are
single-model lapses on a family the paper does state. Because the labels come from the code, a paper
that reported a different family than its script runs would also count as a miss in this arm. That
does not occur in the 37 rows: no model reports a stated family that contradicts the code's for the
labeled analysis. The one apparent case is a Haiku row that records an appendix ordinal-probit
specification of the column instead of the main-text regression the label tracks, a seemingly unrelated regression (SUR)
fit. Adding code raises all three to 34 of 37, one short of the ceiling, and Haiku from 23 to 25.
The one remaining row differs by model. For Opus and Fable it is a control case whose code fits its
zero-inflated models on a lead of the outcome, the next period's value. The models attribute every
zero-inflated fit to that lead, including the one the code also runs on the recorded column. That is
an attribution error on the code's own terms, and it leaves the recorded column with only its
negative-binomial robustness fit. For
Sonnet it is a binary marriage-status outcome fitted inside a code loop, which it does not link. An earlier two-step pilot, a distilling call followed by a resolving call with the same model and
inputs, gave Haiku 11 of 37 on text and metadata alone. The one-call prompt lifts that to 23, and to
25 with code. Single-run differences of one or two rows do not establish a ranking. The
released raw outputs, prompts, scorer, and per-case scores make each row inspectable. The grid
measures extraction under its column-alias and estimator-family rules; it does not score the role
labels or the scientific reading of each recovered link.

\paragraph{Proposed ingest flow}
At ingest, a repository could resolve paper--deposit links, extract analysis records when the needed
sources are available, and materialize alerts over the profile of Section~\ref{sec:profile}.
Deterministic identities permit repeatable imports of the same evidence. Changed extractions create
new assertion versions, and disagreements remain reviewable. Source, generation method, and review
scope stay independent fields, not one combined provenance field. This flow is a design: the example
validates one small export and query fragment, and fingerprint coverage beyond the testbed is
unmeasured.

\section{Discussion}
\label{sec:discussion}

\paragraph{What the contribution enables}
The reusable unit is an assertion that a reported analysis uses a particular versioned variable,
with its role, evidence, and review history. A screening rule is one consumer of those assertions.
Retrieval is another. A researcher who wants to test an estimator on real data rather than on a
stock benchmark needs deposited datasets with a particular kind of variable: a count outcome with a
few hundred observations, a bounded share, an ordinal item. To our knowledge, no index offers that today. The census join
answers the question directly. The 1{,}440 deposits with a matched outcome are a search result before
they are a flag. So are the clear-tier linear fits flagged within them outside the testbed, 758 proportion and 178
count flag records, which collapse to 838 distinct outcome questions. The role and review predicates then let
the query ask for headline analyses or reviewed mappings only.
The example export also retrieves the robustness refit and the balance analysis, which produce no alert. A repository can
therefore expose an analysis inventory and a review queue over the same metadata, selecting
headline, side-analysis, or reviewed views as needed.

The acquisition results show where such a service would lose coverage: only 1,440 of 4,868 deposits
have an extracted outcome matched to a profiled column, and the benchmark traces most misses to
command recovery, outcomes constructed or renamed in code, or missing profiles. The fingerprint
experiment tests another route to analysis evidence, and the export supplies the scope and lineage
structures needed to represent what it returns. Each component has its own evaluation; their joint
coverage in a deployed ingest service remains to be measured.

A flagged combination keeps its value when review changes its interpretation: the logged
battle-deaths response needs a link to its raw count input without losing its own identity, and the
hate-crime example needs the paper's robustness analyses as context. The profile records those
distinctions next to the original extraction, so corrections improve later queries without erasing the
evidence that led to review.

\paragraph{Limitations}
A single-column profile cannot establish conditional-model validity or recover dependence
structure, and an inferred support class does not uniquely identify a substantive variable;
transformations, missing values, and the estimation sample can all change which profile a check
should use. The census refits approximate recognizable specifications rather than reproducing full
workflows. The benchmark was selected around the implemented linear-on-non-continuous rule, so its
saturated score measures consistency with those labels and nothing more. Precision is 55\% on a
roughly balanced type sample, 62\% on proportions and 49\% on counts, after development on the same
corpus; it is neither a prevalence-weighted archive estimate nor a held-out validation, and agreement
between two AI-assisted adjudication sessions is not independent scientific endorsement.

Census coverage stops where the harvest stops: public tab or csv data under the size cap, scripts
within the first eight per deposit, and command forms the keyword mapping recognizes. Coverage
figures are therefore lower bounds on the regressions and outcome links the deposits contain. The control set
is small and mixes deposit-level flag absence with analyses that can coexist in the same deposit. The
fingerprint grid uses a small, selected testbed and single generations, and its role labels are
extracted but not validated.
The application profile is a proposal with a tested example, not a production integration; its local
identifiers and example vocabulary need repository governance before long-term deployment.

\paragraph{Repository adoption}
A deployment would need repository-managed identifiers, a published and versioned extension
vocabulary, and policies defining who can review each kind of assertion. Acquisition could then
proceed incrementally as files, code, and paper text become available. The next evaluation should
measure coverage of analysis--variable links and the effort required to review or correct them
across deposits, including transformed responses and restricted samples. Those measures would test
the service the profile enables: how much usable analysis metadata a repository can acquire and
maintain.

\paragraph{Relation to a broader program}
A domain-stripped fingerprint places papers in one substrate by what they compute~\citep{p1,p3}; the
profile gives datasets a place in it through their variable-level metadata, with the analysis
fingerprint as the paper-side bridge. Two ends remain: scaling the fingerprint layer beyond the
testbed, and the dual of the screening map, which estimator families a neighboring field already uses
for a given outcome class, a directed cross-field suggestion once an at-scale map exists.

\section{Conclusion}
\label{sec:conclusion}

Reading the data back requires a persistent link from a reported analysis to the versioned variables
it uses. The application profile specifies that link together with empirical profiles, analysis
roles, provenance, and separate review decisions. Its worked export of one replication package
(477 triples) illustrates the profile, passing core and extension checks and four queries; in that case the flagged headline
regression is retrieved together with the paper's own quasi-Poisson robustness refit of the same
outcome, which is what turns a bare flag into a scoped review candidate. The code-only census over 4{,}868 datasets quantifies the coverage
and yield of one acquisition route: outcome links in 1{,}440 deposits, 965 flagged, 14 of the 21
in-census benchmark mismatch deposits recovered, and 55\% adjudicated precision on the sampled clear-tier
questions. From text and metadata, the fingerprint experiment recovers both the outcome link and the
estimator family for 23 to 28 of 37 testbed cases; with code added, for 25 to 34. Of the 37, 35
have an outcome column the metadata can carry.

Model-choice screening shows how these relationships can feed an inspectable review queue; the same
representation answers retrieval by variable use, analysis role, or review state.

The natural place for the profile is deposit time. The depositor uploads, the ingester profiles the
variables and drafts the analysis records, and the depositor reviews the variables, roles, and fitted
models before publication. That should be less work than writing the metadata from nothing, and it is done
by the person who knows the analysis. Because review decisions are stored apart from the generated
assertions, the automatic ingest stays on record next to the corrected version. Over many deposits
those pairs would show where the ingest tends to go wrong, which is the evidence needed to improve it.

What the paper contributes is to make those relationships explicit, testable, and reusable as
repository metadata. That metadata is data in its own right. Once every deposit carries the profile,
the archive becomes one table that can be queried and joined across deposits, by variable kind,
sample size, analysis role, estimator family or review state, by a person or by a program. Finding
the deposited datasets on which a method should be compared, or every reviewed headline analysis that
models a bounded share, is then a query rather than a literature search.

\backmatter

\section*{Statements and Declarations}

\bmhead{Funding}
No external funding was received for this work; it was carried out as part of the author's position at
LIBIS, KU Leuven.

\bmhead{Competing interests}
The author declares no competing interests.

\bmhead{Ethics approval}
Not applicable. The study uses only publicly deposited replication data and code and involves no human
participants. Deposits and papers are cited by their public DOIs as scholarly records. The screen's
flags are review candidates for metadata mappings and model-choice questions; none is a finding that
a published analysis is incorrect, and the benchmark and released materials record deposit completeness
and screening output, not assessments of individual studies.

\bmhead{Consent to participate}
Not applicable.

\bmhead{Consent for publication}
Not applicable.

\bmhead{Author contributions}
Eryk Kulikowski is the sole author: he conceived and designed the study, built and verified the
testbed, ran the experiments, and wrote and revised the manuscript.

\bmhead{Data and code availability}
The accompanying reproduction package contains the extended generator, the 40-case testbed with
DOIs and evidence notes, the census flags and refit summaries, and all 160 adjudication
verdict records. It also includes the analysis-fingerprint prompts, metadata inputs, raw outputs and
scores (\texttt{text\_route\_pilot/}), and the application-profile example,
queries, validation shapes and regression checks (\texttt{metadata\_profile/}, \texttt{tests/}).
The driver \texttt{reproduce.py} regenerates the reported aggregates and checks the example's
interchange and query results; raw-data checks are identified separately when the census pool is
absent. The source repository is \url{https://github.com/ErykKul/reading-the-data-back}.
Deposited data, code, and paper texts are not redistributed; fetch scripts and public DOIs identify
the sources. A frozen copy will be archived in the KU Leuven Research Data Repository (RDR) with a
DOI at publication.

\bmhead{AI use disclosure}
Claude (Anthropic) was used as coding and drafting assistant. The author conceived and
directed the research, verified all claims, numbers, and citations, and takes full responsibility for
the final text.

\end{document}